\documentclass[aps,prl,twocolumn,superscriptaddress]{revtex4-2}

\usepackage{graphicx}
\usepackage{epsfig}
\usepackage{dcolumn}
\usepackage{bm}
\usepackage{overpic}
\usepackage{color}
\usepackage{multirow}
\usepackage{subfigure}
\usepackage{lineno}
\usepackage{amsmath}
\usepackage{url}

\begin{document}

\title{How to determine  nucleon polarization at existing collider experiments?}

\author{Yu-Tie Liang}
\email[]{liangyt@impcas.ac.cn}
\affiliation{Institute of Modern Physics, Chinese Academy of Sciences, Lanzhou 730000, People's Republic of China}
\affiliation{University of Chinese Academy of Sciences, Beijing 100049, People's Republic of China}

\author{Xiao-Rong Lv}
\affiliation{Institute of Modern Physics, Chinese Academy of Sciences, Lanzhou 730000, People's Republic of China}

\author{Andrzej Kupsc}
\email[]{andrzej.kupsc@physics.uu.se}
\affiliation{Uppsala University, Box 516, SE-75120 Uppsala, Sweden}
\affiliation{National Centre for Nuclear Research,  Pasteura 7, 02-093 Warsaw, Poland}

\author{Boxing Gou}
\affiliation{Institute of Modern Physics, Chinese Academy of Sciences, Lanzhou 730000, People's Republic of China}
\affiliation{University of Chinese Academy of Sciences, Beijing 100049, People's Republic of China}

\author{Hai-Bo Li}
\email[]{lihb@ihep.ac.cn}
\affiliation{Institute of 
High Energy Physics, Chinese Academy of Sciences,  Beijing 100049, People's Republic of China}
\affiliation{University of Chinese Academy of Sciences, Beijing 100049, People's Republic of China}

\date{\today}

\begin{abstract}

We propose a novel approach to measure spin polarization of nucleons produced in electron--positron collisions. Using existing tracking devices and supporting structure material, general-purpose spectrometers can be utilized as a large-acceptance polarimeter without hardware upgrade. With the proposed approach, the spin polarization of nucleons can be revealed, providing a complementary and accurate description of the final-state particles.  This could have far-reaching implications, such as enabling the complete determination of the time-like electromagnetic form factors of nucleons. 

\end{abstract}

\maketitle


Spin is a fundamental property of atoms and subatomic particles. It affects their interactions and is essential for explaining the structure of nuclei, atoms and the properties of materials. The study of spin has led to many important discoveries and continues to be a very important area of research, e.g., precision determination of muon anomalous magnetic moment as a test of the Standard Model of elementary particles~\cite{Muong-2:2021ojo,Muong-2:2023cdq}.

In nuclear and particle physics, many experiments are performed using spin-polarized beams or targets,
where the polarization can be controlled~\cite{BARBER1995436} and determined using dedicated devices~\cite{NakagawaAIP, PhysRevD.79.094014, ComptonHERA, ComptonJLab, Woods:1996nz, MottJLab2020}. This ensures a complete and accurate description of the initial state of the reactions. To measure the polarization of the final-state particles, a variety of dedicated detectors are built that act as secondary polarimeters. In the case of decaying particles, the polarization can be determined from distributions of the daughter particles. In particular, complete determination of the polarization is possible when parity-violating weak decays are involved. Due to this property, polarization of hyperons is investigated in various reactions, such as at electron--positron collision~\cite{BESIII:2018cnd, PhysRevLett.122.042001}, proton--proton or proton--nucleus~\cite{PhysRevLett.36.1113, PhysRevLett.67.1193} and heavy ion collisions~\cite{PhysRevC.104.L061901}. 
For the final states involving protons, the spin information is normally not measured since it would require an external polarimeter based on secondary interactions. Many such devices have restricted angular acceptance or cannot be accommodated modern detectors. This has made the information on the final state incomplete, limiting the experimental studies in the field.

In this Letter, we propose a novel approach that uses information collected by the standard detectors to measure the spin polarization of the nucleons. Using the tracking detector and certain supporting material, most general-purpose spectrometers have a potential to be large-acceptance polarimeters without hardware modification. Here, we evaluate the feasibility of such polarization measurements using as an example the currently operating electron--positron collider experiment BESIII~\cite{ABLIKIM2010345}. A similar method can be easily adopted in various experiments at electron--position colliders, pp/pA and lepton proton scattering machines with low track multiplicity and outgoing nucleons with laboratory momenta up to a few GeV. 
For planned or to be upgraded detectors, we propose minor hardware modifications to increase efficiency of the polarization measurement without a significant impact on the detector performance.

The BESIII detector records symmetric $e^+e^-$ collisions provided by the BEPCII storage ring operating
in the center-of-mass (c.m.) energy range $\sqrt{s}=1.8-5.0$~GeV. 
BESIII has collected $10^{10}$ $J/\psi$ events~\cite{BESIII:2021cxx} including processes such as $J/\psi\to p\bar p$ and $J/\psi\to \Lambda\bar \Lambda$ with branching fractions $2.12(3)\times10^{-3}$ and $1.88(8)\times10^{-3}$, respectively ~\cite{PhysRevD.110.030001}.
The BESIII detector is described in detail in ~\cite{ABLIKIM2010345}. The cylindrical core of the BESIII detector covers polar angles range $35^\circ-145^\circ $  and consists of a helium-based multilayer drift chamber~(MDC), a plastic scintillator time-of-flight system~(TOF), and a CsI(Tl) electromagnetic calorimeter~(EMC), which are all enclosed in a superconducting solenoidal magnet providing a 1.0~T  magnetic field. The charged-particle momentum resolution at $1~{\rm GeV}/c$ is $0.5\%$.

A measurement of a proton polarization usually utilizes the spin-dependent cross section for the proton--proton  ($pp$) or proton--carbon ($p\textrm{C}$) elastic scattering given their large analyzing powers. In the case of unpolarized target, the differential cross section for a given c.m.  polar angle $\theta$ is expressed as~\cite{PhysRevLett.90.142301, Bystricky:1976jr}: 
\begin{equation}
\frac{d\sigma}{d\phi d\cos\theta} =  \frac{1}{2\pi}\frac{d\sigma_0}{ d\cos\theta}\left[1+P_yA_N\!(\theta)\cos\phi\right]
\end{equation}
Here, $\sigma_0$ indicates the unpolarized cross section and $\phi$ is the azimuthal angle and $P_y$ is the transverse polarization of the proton to be determined. $A_N$ is the analyzing power.
In collider experiments, such as BESIII, the produced nucleons usually have momenta ranging from a few hundred ${\rm MeV}/c$ to a few ${\rm GeV}/c$. The analyzing powers for the reactions $pp$ and $p\textrm{C}$ have been determined with high precision in this energy region~\cite{PhysRev.148.1289,PhysRev.163.1470, PhysRev.95.1348, ALBROW1970445, PhysRevC.24.1778, vonPrzewoski:1998ye, PhysRevLett.41.384, PhysRevD.21.580, PhysRevD.40.35, PhysRev.105.288, GREENIAUS1979308}  and the values are large --- up to 60\%~\cite{PhysRevLett.41.384}.

Hydrogen and carbon can be found in many construction elements in the detectors. For example, in the BESIII detector, a layer of cooling fluid of high purity mineral oil with the thickness of 0.8~mm circulates between inner and outer beryllium shells of the beam pipe. The inner radius of the oil layer is 32.3~mm~\cite{ABLIKIM2010345, PhysRevLett.127.012003}. The mineral oil is mainly composed of carbon and hydrogen, which are the perfect target material. The tracking detector has an inner tube wall made of carbon fiber with a thickness of 1.2 mm, which is also a good target candidate. These supporting components of the detector were originally undesired and were optimized to be as thin as possible to improve detector performance. Here, we utilize these supporting layers to prove the feasibility of measuring the polarization of final-state nucleons at the high luminosity colliders. 

\begin{figure}
\includegraphics[width=0.42\textwidth]{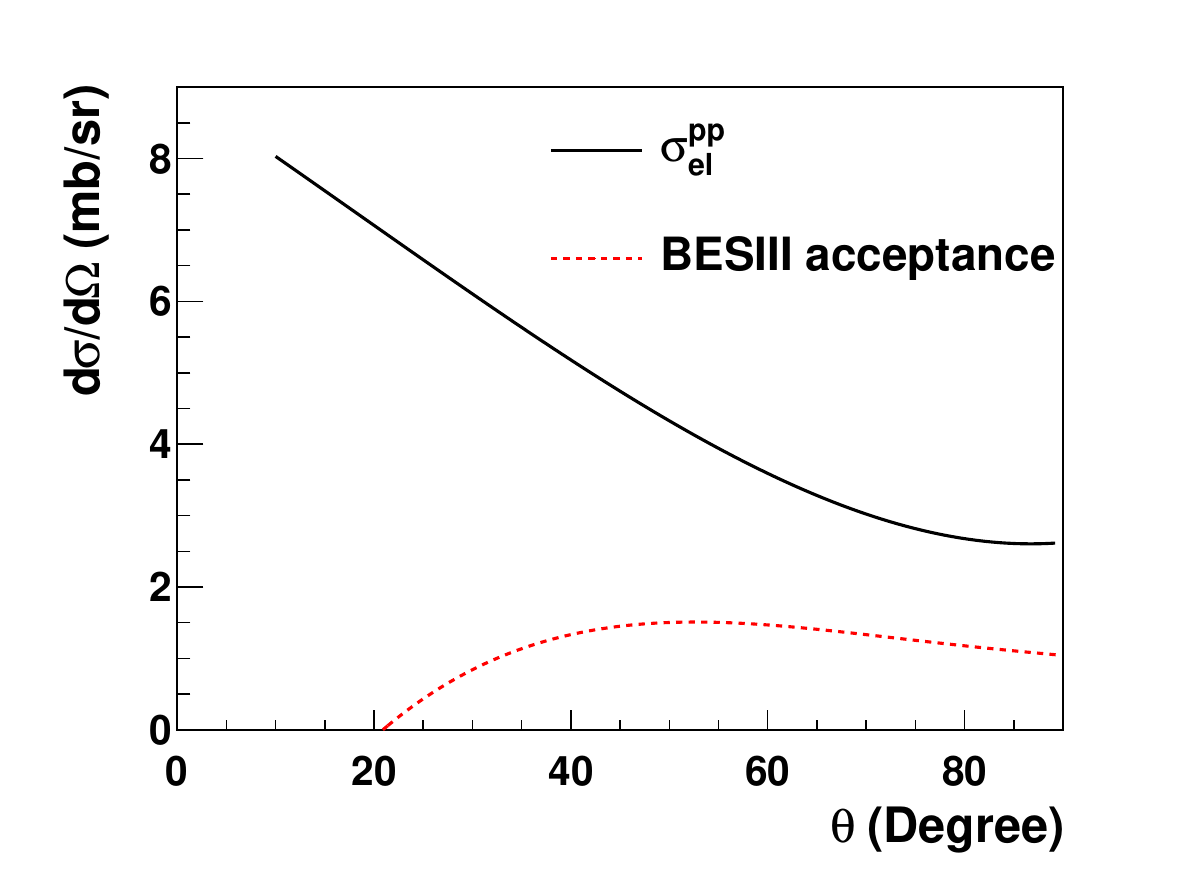}
\includegraphics[width=0.42\textwidth]{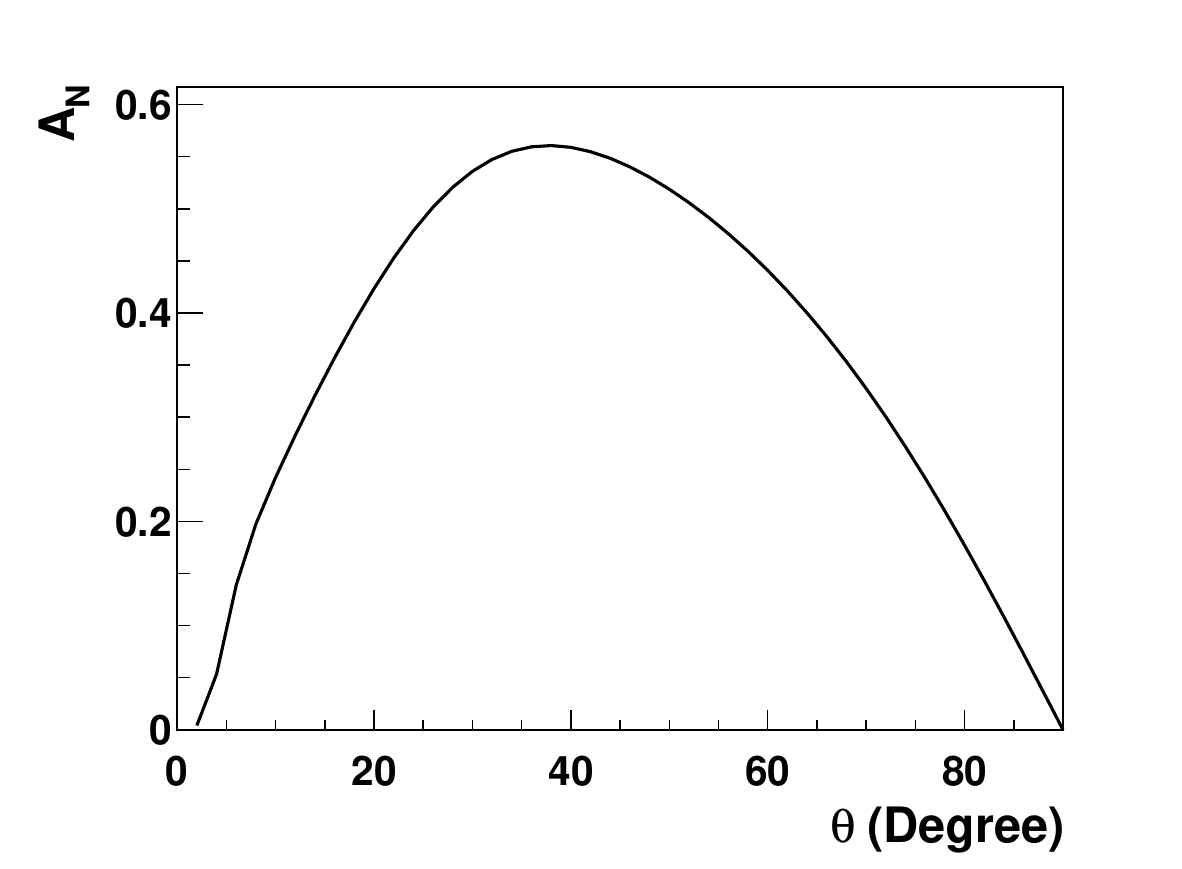}
\caption{\label{fig:example_jpsi_ppbar} {(Upper plot) The $pp$ elastic differential cross section  at proton momentum of 1.23 GeV with a fast simulation including the BESIII detector acceptance. (Bottom plot) The proton analyzing power for this process~\cite{GWDAC}.}
}
\end{figure}

We take as example (anti)protons from the $e^+e^-\to J/\psi\to p\bar p$ process, 
where a significant transverse proton spin polarization is possible such as observed in  the $J/\psi$ decays into a hyperon--antihyperon pairs~\cite{BESIII:2018cnd,BESIII:2021ypr}.
In this process the (anti)proton momenta are $1.23$~GeV/$c$, implying the c.m. energy of the elastic $pp$ scattering with the detector material is 2.16 GeV. The differential cross section of elastic scattering~\cite{GWDAC} is given by the black solid curve in the upper plot of Fig.~\ref{fig:example_jpsi_ppbar}, while the cross section after taking into account \mbox{BESIII} acceptance is given by the red dashed curve.  The angular dependence of the analyzing power at this energy is shown in the bottom plot of Fig.~\ref{fig:example_jpsi_ppbar}. The average analyzing power is 0.4.

\begin{figure*}
\includegraphics[width=0.8\textwidth]{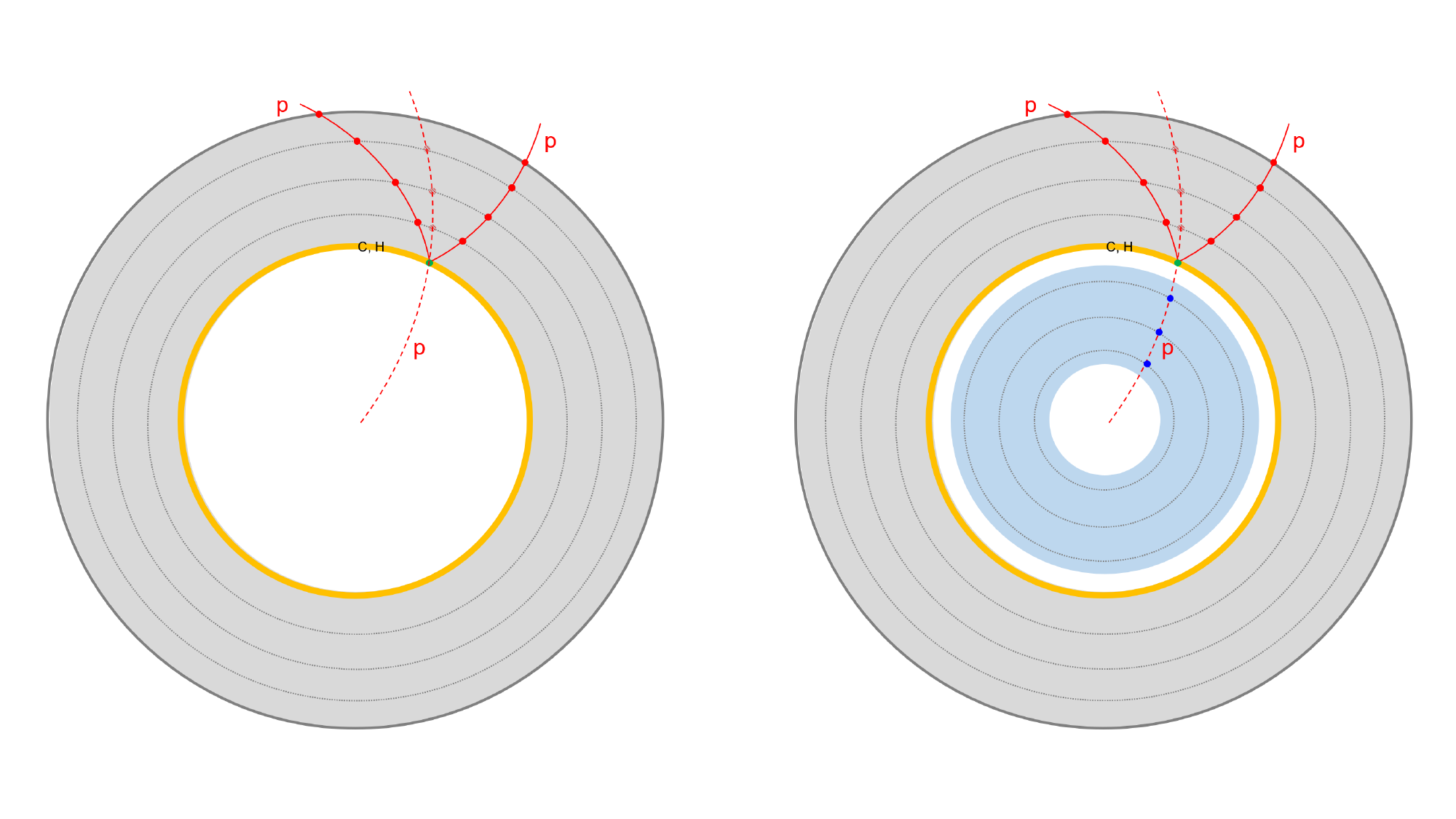}
\caption{\label{fig:method} Illustration of $pp$ elastic scattering event  at BESIII detector. (Left) In case the scattering layer is downstream of tracking detector, a full event could be reconstructed to deduce the original proton (dashed line). (Right) In case the scattering layer is in between the tracking detector, both initial proton and two final state protons are measured. Non-cylindrical detectors is also applicable.}
\end{figure*}

An elastic scattering event $pp$ is illustrated in Fig.~\ref{fig:method}. In the right plot, the original proton and the final-state protons (recoil and scattered proton) are all detected. Complete information of $pp\rightarrow pp$ is available, from which polarization can be extracted. The plot on the left shows a setup with the secondary scattering layer upstream of the tracking detector. The two final-state protons are detected, while the original proton is not. This is the case of the BESIII experiment, where the scattering layers are the beam pipe or the inner MDC wall. Here, the complete reconstruction of an event can be performed due to the low multiplicity of the final states. In particular, the projectile proton momentum can be deduced from the kinematical constraints. 

\begin{figure}
\includegraphics[width=0.42\textwidth]{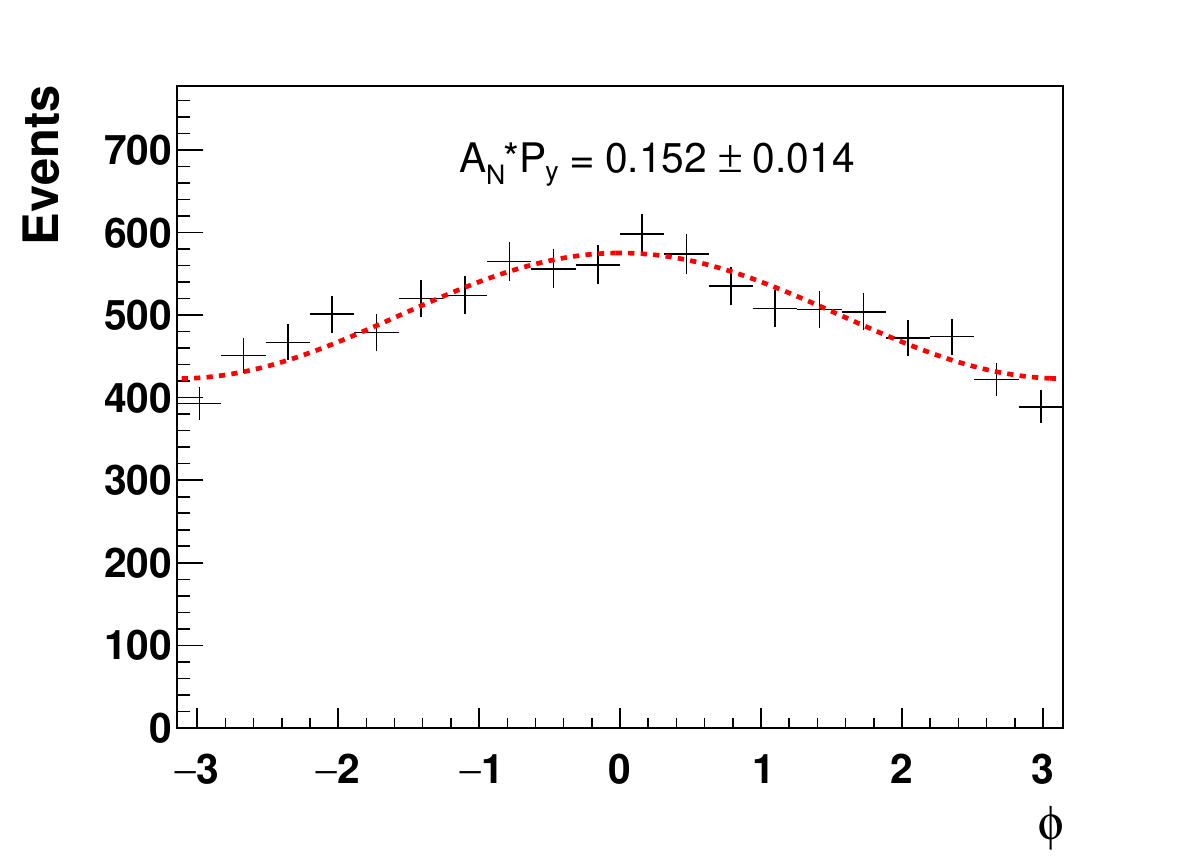}
\caption{\label{fig:err_band} Single spin asymmetry ($A_{SS}$) extracted from the $\phi$ distribution. In MC, an input transverse polarization of 0.3 and an analyzing power of 0.5 is assumed. 
}
\end{figure}

The method utilizes a tiny layer of hydrogen- or carbon-rich material in the existing detector, so the probability of $pp$ scattering is small. However, it could be compensated at high luminosity machines, such as the BESIII and the future STCF~\cite{Achasov:2023gey}. Taking BESIII as an example, several millions of (anti)nucleons from $J/\psi$ have been collected. The probability of the proton elastic scattering off a proton in the oil layer of 0.8 mm is roughly $1.2\times10^{-4}$, assuming a 20 mb $pp$ elastic scattering cross section in acceptance of the BESIII~\cite{PhysRevD.110.030001}. Taking into account detector efficiency, we can expect to have an order of $N=10^{3}$  $pp$  elastic scattering events to measure spin polarization. Fig.\ref{fig:err_band} shows the extracted single spin asymmetry with a MC dataset of 10000 signal events. In the MC, an input transverse polarization of 0.3 and a $pp$ analyzing power $A_N=0.5$ are assumed. 

The statistical uncertainty of the product $A=P_yA_N$ determination using maximum log-likelihood fit is:
\begin{align}
 \sigma(A)
 &\approx\sqrt{\frac{2}{N}}\ .  \nonumber
\end{align}
Therefore, for $10^{3}$ $pp$ elastic secondary scattering events, as expected from the $J/\psi\to p\bar p$ process at BESIII, $\sigma(A)=0.04$ and we can expect uncertainty of 0.1 for the proton polarization in the decay.

Therefore we have got promising results while considering only the $pp$ elastic scattering signal. The $p\textrm{C}$ scattering events have the potential to further improve the precision of the polarization measurement. In the $p\textrm{C}$ case, the identification of the carbon using energy deposition or time-of-flight is nontrivial and has to be investigated; alternatively, if the original proton is measured,the carbon could be left undetected. This method can be used to measure antiproton, neutron, and antineutron, as long as the analyzing powers of the corresponding processes are determined~\cite{KUNNE1988557}. 

For the existing detectors, a single element target dedicated as a polarimeter is hard to find. In BESIII, the target is mineral oil, a mixture of hydrogen and carbon. The selected $pp \rightarrow pp$ signal might be from a collision with a quasifree proton in the Carbon nucleus. In order to suppress this background, both final-state protons shall be reconstructed.  Combined with the original proton, a four-momentum constraint can be applied to suppress the quasi-free collisions. An MC study indicates that the background level can be reduced to below 3\%. In case of a pure hydrogen target, the signal reconstruction will be simpler. For example, detection of only one final-state proton would improve an overall efficiency.
The detection efficiency can also be improved if the thickness of the scattering layer is enlarged. An optimization was performed to study the influence of the increased material budget on the resolution. In case the thickness is enlarged by an order of magnitude (from 0.2\% to 2\% X$_0$)~\cite{ABLIKIM2010345},  the detector efficiency and the momentum resolutions of the final-state proton decrease by less than 1\%. This exercise indicates a minor impact on the normal detector performance, whereas the efficiency of polarization measurement is increased by one order of magnitude.

To validate the polarization measurements, protons with known polarizations are needed.
For this purpose, protons from hyperon decays can be used. For example, the proton polarization from  unpolarized $\Lambda$ decay into $p\pi^-$ is $\alpha_\Lambda=0.75$~\cite{PhysRevLett.129.131801, PhysRevD.110.030001} and directed along the proton momentum in the $\Lambda$ rest frame. In the $J/\psi\to\Lambda\bar\Lambda$ reaction at BESIII, the protons acquire transverse polarization with respect to the direction of the proton in the detector reference frame. 

\begin{figure}
\includegraphics[width=0.45\textwidth]{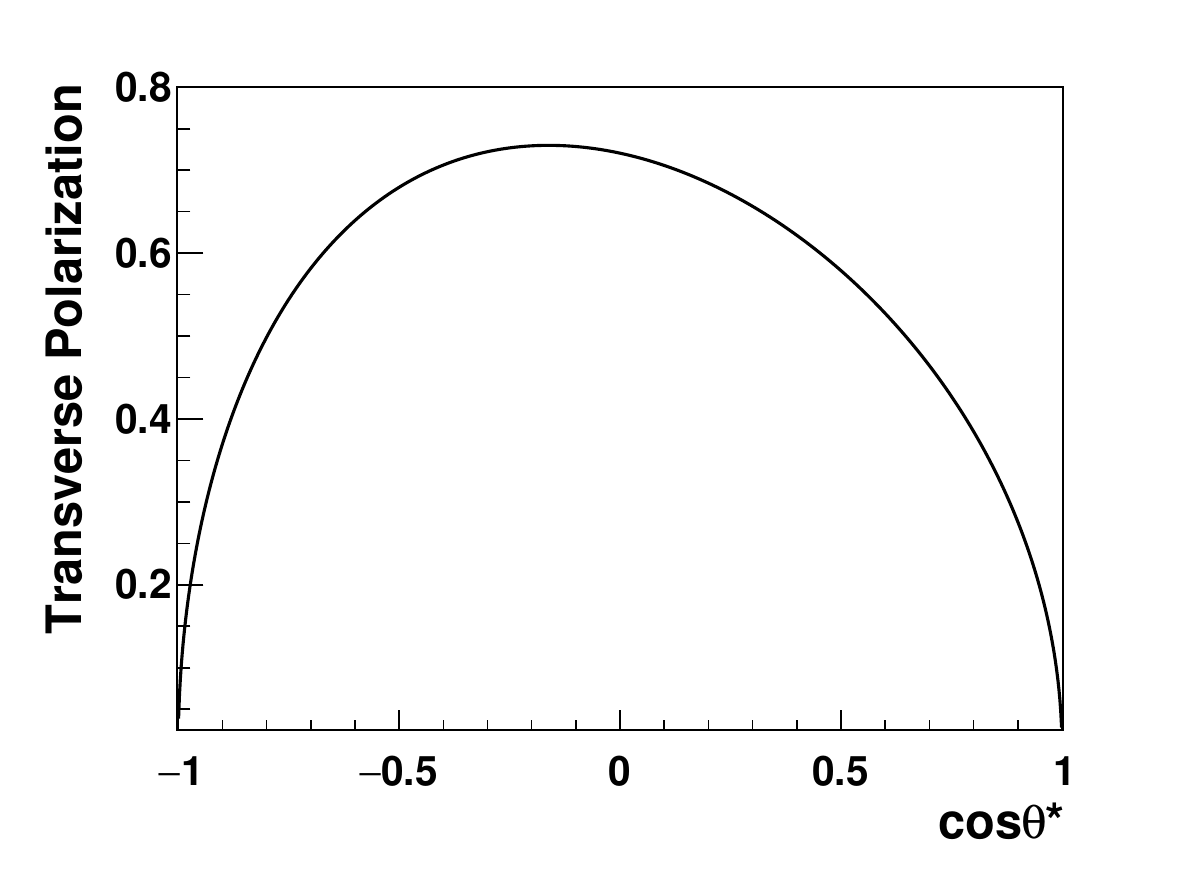}
\includegraphics[width=0.45\textwidth]{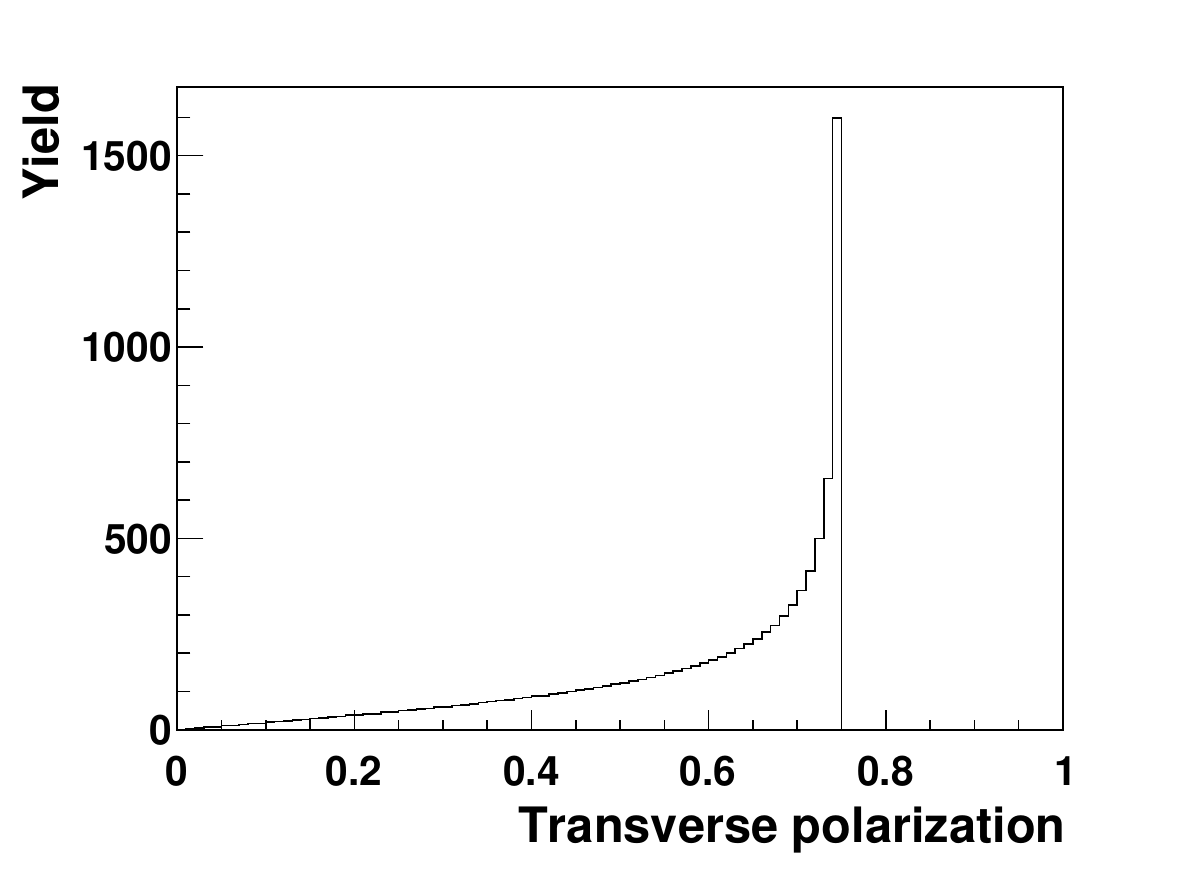}
\caption{\label{fig:pol_p_from_lambda} (Upper plot)Transverse polarization of proton from unpolarized $\Lambda$ decay as a function of $\mathrm{cos\theta^*}$ of proton momentum in $\Lambda$ rest frame. (Bottom plot) Distribution of transverse polarization of proton from unpolarized $\Lambda$ decay.}
\end{figure}

The protons emitted from the decay of $\Lambda$ with 1 GeV/$c$ lab momentum have transverse polarization distribution shown in Fig.~\ref{fig:pol_p_from_lambda} with an average value of 0.59. In the upper plot, $\theta^*$ is the angle between the proton momentum in the $\Lambda$ rest frame and the momentum $\Lambda$ in the laboratory frame. This is a well-known way to produce polarized (anti)proton beams that was used at Fermilab in the 1990s~\cite{GROSNICK1990269}. The polarization is nearly independent of the $\Lambda$ momentum, ensuring high and controlled proton polarization.

\begin{figure}
\includegraphics[width=0.45\textwidth]{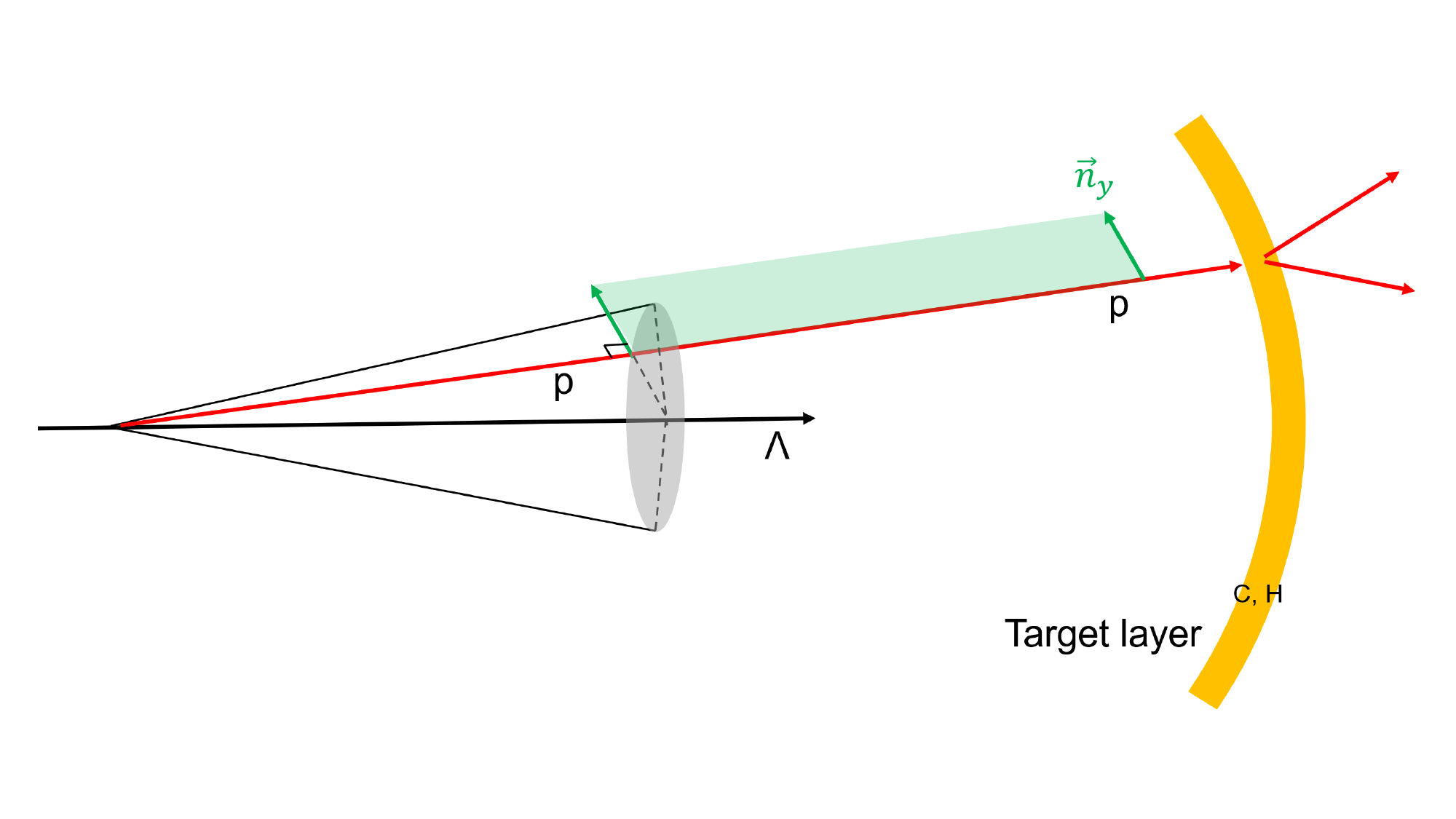}
\caption{\label{fig:p_from_lambda} Transverse polarization axis of proton from $\Lambda$ decay. Red lines with arrow indicate the momentum of daughter proton. The daughter pion is not shown. Green lines with arrow indicate the transverse polarization axis of proton.}
\end{figure}

Most existing or future facilities can use this method to prepare polarized protons as long as there is enough energy to produce hyperons. Fig.~\ref{fig:p_from_lambda} shows the transverse polarization axis of the proton from the decay of $\Lambda$. 
This source of polarized (anti)nucleons could be also used to determine the analyzing powers for processes where they are not precisely known, such as $\bar{p}p$, $np$, or $\bar{n}p$.

The method can be adopted for various reactions. For an experiment with low multiplicity similar to the BESIII, which is normally the case in $e^+e^-$/$ep$/$pp$ collisions, it is not required that the projectile proton is detected. 
For experiments with high multiplicity, such as in heavy-ion collisions,  the projectile proton should be detected to extract the full information on the $pp$ scattering. This indicates that the polarimeter target 
should be placed in between the tracking detector layers, allowing the detection of both the projectile and the two outgoing protons. 

Many applications of the nucleon polarimeter can be foreseen in nuclear and particle experiments.  The polarization of the final-state proton (or antiproton) in $e^+e^-\rightarrow p\bar{p}$ is directly related to the relative phase between the electric and magnetic form factors of the proton. Therefore, using our method, the time-like form factors could be completely determined for the first time. This approach is also applicable to neutron form factors, but the detection of neutrons is more challenging. 

Future electron ion colliders (EIC and EicC)~\cite{AbdulKhalek:2021gbh, Anderle:2021wcy} aim to perform multidimensional imaging of the internal structure of the nucleon. With the nucleon polarization determined, more spin-related observables can be extracted to benefit both the experimental and theoretical frameworks.  

In the $pp$/$p\textrm{A}$ collision, many measurements of hyperon polarization have been performed to study the production mechanism. Whence the polarization of the nucleons is determined, it is possible to analyze how it arises from various sources. In heavy-ion collisions, the global polarization could be studied in similar way as for $\Lambda$s to bring new insights into properties of the extreme media created in such conditions~\cite{PhysRevC.104.L061901}.

\begin{acknowledgments}
 This work is supported in part by the National Key Research and Development Program of China under Contrac No. 2023YFA1606800; The National Natural Science Foundation of China (NSFC) under Contracts No. 11975278; Swedish Research Council under grant No. 2021-04567; {Polish National Scien\-ce Centre through the grants 2019/35/O/ST2/02907 and 2024/53/B/ST2/00975}. Y.T.L thanks Yi Yin, Changzheng Yuan, Nu Xu for helpful discussions.
\end{acknowledgments}

\bibliography{polarimeter}

\end{document}